\documentclass[prd,twocolumn,floatfix]{revtex4}
\usepackage[dvips]{graphics}
\usepackage{amsmath}

\newcommand{\figsize}{}

\input prepictex
\input pictex
\input postpictex
\newdimen\tdim
\def\stpltsmbl{\setplotsymbol ({\small .})}
\def\bsmbl{\setplotsymbol ({\Huge .})}
\def\tarrow{\arrow <5\tdim> [.3,.6]}

\newcommand{\moose}[3]{\startrotation by .383 .924
#1
\stoprotation
\startrotation by .707 .707
#2
\stoprotation
\startrotation by .924 .383
#1
\stoprotation
#2
\startrotation by .924 -.383
#1
\stoprotation
\startrotation by .707 -.707
#2
\stoprotation
\startrotation by .383 -.924
#1
\stoprotation
\startrotation by 0 -1
#2
\stoprotation
\startrotation by -.383 -.924
#1
\stoprotation
\startrotation by -.707 -.707
#2
\stoprotation
\startrotation by -.924 -.383
#1
\stoprotation
\startrotation by -1 0
#2
\stoprotation
\startrotation by -.924 .383
#1
\stoprotation
\startrotation by -.707 .707
#2
\stoprotation
\startrotation by -.383 .924
#1
\stoprotation
#3}

\makeatletter
\renewcommand{\@makecaption}[2]{
  \vskip\abovecaptionskip
  \sbox\@tempboxa{\small\sf #1: #2}%
  \ifdim \wd\@tempboxa >\hsize
  \small\sf #1: #2\par
  \else
    \global \@minipagefalse
    \hb@xt@\hsize{\hfil\box\@tempboxa\hfil}%
  \fi
  \vskip\belowcaptionskip}
\makeatother

\providecommand{\abs}[1]{\lvert#1\rvert}

\begin{document}
\preprint{HUTP-01/A040}
\preprint{BUHEP-01-21}
\preprint{LBNL-48727}

\title{Twisted supersymmetry and the topology of theory space}

\author{Nima \surname{Arkani-Hamed}}
\email{arkani@bose.harvard.edu}
\altaffiliation[Permanent Address: ]{Department of Physics, UC
  Berkeley, Berkeley CA 94720}
\author{Andrew G. Cohen}
\email{cohen@andy.bu.edu}
\altaffiliation[Permanent Address: ]{Physics Department, Boston
  University, Boston MA02215}
\author{Howard Georgi}
\email{georgi@physics.harvard.edu}
\affiliation{Lyman  Laboratory of Physics, Harvard University,
  Cambridge MA 02138}

\date{September 2001}

\begin{abstract}
  We present examples of four dimensional, non-supersymmetric field
  theories in which ultraviolet supersymmetry breaking effects, such as
  bose-fermi splittings and the vacuum energy, are suppressed by
  $(\alpha/4 \pi)^{N}$, where $\alpha$ is a weak coupling factor and
  $N$ can be made arbitrarily large.  The particle content and
  interactions of these models are conveniently represented by a graph
  with sites and links, describing the gauge theory space structure.
  While the theories are supersymmetric ``locally'' in theory space,
  supersymmetry can be explicitly broken by topological obstructions.
\end{abstract}


\maketitle

\section{Introduction}

We do not know how Nature chooses the pattern of gauge symmetries and
matter fields that describes the relativistic quantum mechanics of our
low energy world. For a variety of reasons \cite{Georgi:1986hf,
  Douglas:1996sw, Kachru:1998ys, Bershadsky:1998mb}, it is interesting
to consider a class of four-dimensional gauge theories described by
graphs of ``sites'' and ``links''. A graph consists of a number of
points, or ``sites'', $(i)$, for $i = 1, \dots,N_{\text{tot}}$. Some
pairs of sites are connected by one or more lines, which are called
links. The links between the pair of connected sites $(i)$ and $(j)$
are denoted by $(i,j)_\alpha$, for $\alpha = 1, \dots, n_{ij}$.

We associate a field theory with a graph by assigning gauge symmetries
and matter fields as follows. A gauge group $G_i$ is associated with
each site $(i)$; the full gauge symmetry $G$ of the theory is then $G
= \prod_i G_i$.  We may think of the gauge fields for $G_i$ as
residing on the site $(i)$ in the graph. In addition there may be
additional fields on the site $(i)$ that transform non-trivially under
$G_i$ and trivially under the rest of the gauge symmetry.  Finally,
with each link $(i,j)_\alpha$ in the graph, we associate a matter
field transforming non-trivially under irreducible representations of
$G_i \times G_j$ and trivially under the rest of the gauge group.

All gauge theories can be described in this way with only a single
site associated with the full gauge group. But this picture becomes
nontrivial and interesting when there are several sites.  In this
case, the condition that all matter fields are either site fields or
link fields is a nontrivial constraint on the matter content of the
theory. ``Moose'' or ``Quiver'' models provide simple examples of this
class of theories in which the $G_i$ are unitary groups, and the link
fields transform as bifundamentals under the linked gauge subgroups.

We will refer to the graph associated with a given field theory as the
``theory space'' $T$ of the gauge theory. In previous work
\cite{Arkani-Hamed:2001ca}, we presented examples of theories where
theory space can be transmuted dynamically into field-theoretic
spatial dimensions. Related ideas were presented in
\cite{Hill:2000mu, Cheng:2001vd}. This happens when the link fields
higgs the gauge groups they touch. The effective action in this phase
contains ``hopping'' interactions, allowing field excitations to move
from site to site. The theory space then becomes a picture of
discretized  extra dimensions.  Among other things, this
``deconstruction'' of extra dimensions provides an ultraviolet
completion of higher-dimensional gauge theories
\cite{Arkani-Hamed:2001ca}: at energies far above the higgsing scale
the extra dimensions can melt away into asymptotically free,
perturbative four-dimensional dynamics.

More interestingly \cite{Arkani-Hamed:2001ca}, these constructions
show that the dynamics of four dimensional theories characterized by
theory spaces can be surprisingly rich, with qualitatively new
possibilities beyond those of field-theoretic extra dimensions.  For
example, in \cite{Arkani-Hamed:2001nc}, a new class of realistic
models of electroweak symmetry breaking were constructed, with a
naturally light Higgs field and perturbative new physics at the TeV
scale, but without supersymmetry.  More recently in
\cite{Arkani-Hamed:2001vr}, the successful unification of gauge
couplings in the supersymmetric standard model was accelerated to
occur at energies much lower than the usual grand unification scale.
Other applications have been pursued in \cite{Csaki:2001em,
Cheng:2001an, Cheng:2001qp, Csaki:2001qm,Cheng:2001nh, He:2001fz,
Bander:2001qk,Sfetsos:2001qb, Dai:2001bx, Alishahiha:2001nb}.
 
The sense in which theory space is a generalization of ``geometric''
space is perhaps best appreciated in the context of supersymmetric
models \cite{Arkani-Hamed:2001ca, Csaki:2001em}, which typically have
moduli spaces of vacua. The emergence of extra dimensions from theory
space happens only in certain regions of the moduli space, while in
other regions no geometric interpretation is possible. As an example,
consider an ${\cal N}=1$ $U(1)^N$ supersymmetric gauge theory whose
theory space is shown in fig. 1.
{\figsize%
\begin{figure}[htb]
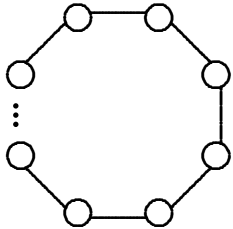

$$\beginpicture
\setcoordinatesystem units <1\tdim,1\tdim>
\stpltsmbl
\moose{\circulararc 360 degrees from 10 80 center at 0 80
}{
\plot -21.38 77.7 21.38 77.7 /
}{
\bsmbl\plotsymbolspacing=7\tdim
\startrotation by 0 1
\plot -7 77.7 7 77.7 /
\stoprotation
\linethickness=0pt
\putrule from 0 -100 to 0 100
\putrule from -105 0 to 105 0}
\endpicture$$
\caption{\figsize\sf\label{fig-1s}Theory space for a $U(1)^N$ 
model.}
\end{figure}%
}
The link $\Phi_i$ between the $i$th and $(i+1)$th site represents a
chiral superfield with charge $(+1,-1)$ under $U(1)_i \times
U(1)_{i+1}$.  This theory has a classical moduli space of vacua with
$\Phi_1 = \dots = \Phi_N = \phi$, characterized by the gauge invariant
operator $(\Phi_1 \cdots \Phi_N)$. For $\phi \neq 0$, the gauge
symmetry is higgsed to the diagonal $U(1)$, with all but one of the
gauge multiplets becoming massive. For simplicity we will take all
gauge couplings equal $g_i=g$.  At energies $E \ll g \phi$ the physics
of the model is the same as that of a five dimensional supersymmetric
$U(1)$ gauge theory compactified on a circle of radius $R \equiv N/(2
\pi g \abs{\phi})$. Here the fifth dimension is discrete, with a
lattice spacing $a \equiv 1/(g \abs{\phi})$.  The mass matrix for
the vector multiplets is
\begin{equation}
  M^2_{ij} = \frac{1}{a^2} \left(2 \delta_{i,j} -
    \delta_{i,j+1} - \delta_{i,j-1}\right)
\end{equation}
where we identify $i$ with $i+N$.
The eigenvalues are 
\begin{equation}
  m_n^2 =  \left(\frac{2}{a}\right)^2 
  \sin^2\left(\frac{na}{2R}\right), 
  \qquad -N/2 < n \leq N/2 
\end{equation}
which reproduces the Kaluza-Klein spectrum of a fifth dimension for
the modes much lighter than $1/a$. Since $\phi$ is a modulus, so is
the radius $R$ of the generated dimension.  Additional superpotential
interactions would stabilize this modulus.

While the physics is that of an extra dimension for $E \ll g \phi$,
for $E \gg g \phi$ this interpretation breaks down and the physics of
the unbroken $U(1)^N$ gauge theory is recovered. In particular, at the
origin of the classical moduli space $\phi = 0$, no extra dimension is
generated at any energy.

This $U(1)^N$ model becomes ill-defined at high energies where the
$U(1)$ gauge couplings are near their Landau poles, but asymptotically
free theories may be constructed as well. For example replace the
$U(1)$ gauge groups by $SU(k)$ and the $\Phi_i$ by bi-fundamental
fields. In this case there is a much larger moduli space of $D$-flat
directions, including the directions where $\Phi_i =
\phi_i\mathbf{1}_k$, characterized by the gauge-invariant operators
$\det \Phi_i$.  In regions of the classical moduli space where the
$\phi_i \neq 0$, geometric space is generated as before. But there are
many regions in classical moduli space with no extra-dimensional
physics, for example along the direction where only $\phi_1 \neq 0$.
As more of the $\phi_i$ are turned on, the full circular geometric
space is built up.  If only $\phi_1,\dots,\phi_{N/2} \neq 0$, half of
the theory space, corresponding to the higgsed links, turns into a
finite interval fifth dimension, while the other half does not.

These examples show that, at least in this class of models, geometric
extra dimensions are not fundamental, but merely an appropriate low
energy description of the physics of the four dimensional gauge theory
on some parts of the classical moduli space. What is more fundamental
is the underlying gauge and matter content of the four dimensional
theory, encoded in the theory space itself.

The structure of field theories characterized by theory spaces is
largely unexplored.  In this paper, we show that theory space can have
interesting topological properties whether or not geometric space
arises.  This allows us to construct non-supersymmetric field theories
where ultraviolet supersymmetry breaking effects can be suppressed by
arbitrarily many perturbative loop factors, allowing for the
generation of large hierarchies of scales in a novel way. On the
portions of the classical moduli space (if any) where geometric extra
dimensions are generated, this way of breaking supersymmetry is
equivalent to the Scherk-Schwartz mechanism \cite{Scherk:1979zr}.
However, some of the most interesting aspects of this way of breaking
supersymmetry occur at the origin of the classical moduli space where no
extra-dimensional Scherk-Schwartz interpretation is possible.

\section{Twisted Supersymmetry}

Let us begin with a simple example by returning to our $U(1)^N$ model.
In addition to the gauge interactions, the bare Lagrangian has the
gauge-Yukawa couplings
\begin{equation}
  \label{eq:2}
  \sum_{i=1}^N   \sqrt{2} \phi^*_i \left( h_i \lambda_i -
    h'_{i-1} \lambda_{i-1}\right) \psi_i +   \text{h.c.}
\end{equation}
where $\phi_i, \psi_i$ are the scalar and fermionic components of
$\Phi_i$, $\lambda_i$ are the gauginos, and we identify the index $0$
with $N$. Supersymmetry requires that these Yukawa couplings are equal
to the gauge couplings: $h_i,h'_i=g_i$. However, we can always
redefine the fields $\phi_i,\psi_i,\lambda_i$ by rephasing them, so
that while the magnitudes of $\abs{h_i}, \abs{h'_i}$ are equal to the
gauge couplings $g_i$, their phases are non-zero. Let us the consider
the more general possibility that the couplings $h_i,h'_i$ are equal
in magnitude to the $g_i$, but with arbitrary phases. By rephasing the
fields we can remove such phases {\em locally} in theory space; for
example we can rephase fields so that $h_1 = g_1, h'_0=g_0$. This
rephasing will in general modify the coefficient of a $\theta$-term in
the Lagrangian, but since such a term is supersymmetric, it
won't concern us here.  However there is one phase that is physical
and cannot be removed by field redefinitions.  Just as the Jarlskog
parameter provides a rephase-invariant measure of CP violation, we can
characterize supersymmetry breaking by the rephase invariant quantity
$J$:
\begin{equation}
  \label{eq:3}
  J \equiv \prod_{i=1}^N h_i^* h'_{i-1} \equiv
  g_1^2\dots g_N^2  V
\end{equation}
where $V = e^{2 i \theta}$ is a phase.  If $\theta$ is non-zero,
supersymmetry is explicitly broken. But this breaking is non-local,
involving {\em all} of the Yukawa interactions. Therefore, any
supersymmetry breaking effect must involve $J$ and will be suppressed
at large $N$. For example if we expand around the vacuum in which
$\phi=0$, the first non-zero contribution to the vacuum energy comes
from the $N+1$ loop diagram of fig. 2 plus its superpartners.
Similarly the leading contribution to scalar masses arises at $N$
loops.  Cancellation between bosons and fermions, which would normally
ensure the absence of any dependence on large momenta, is imperfect
because of the global twist which breaks supersymmetry, proportional
to $(\text{Re} J - g_1^2\dots g_N^2)$. Consequently the vacuum energy
and scalar masses depend sensitively on the details of the UV physics,
which we represent here by a cut-off $M$.

{\figsize%
\begin{figure}[htb]
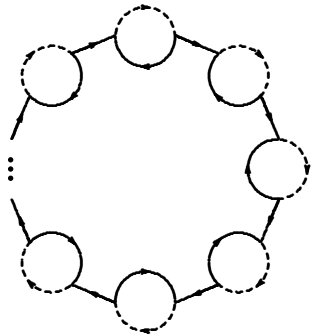

$$\beginpicture
\setcoordinatesystem units <1\tdim,1\tdim>
\stpltsmbl
\moose{\tarrow from -17.5 102.03 to 2 102.03
\plot -17.5 102.03 17.5 102.03 /
}{
\setdashes <4\tdim>
\circulararc 180 degrees from 22.48 100 center at 0 100
\setsolid
\circulararc 180 degrees from -22.48 100 center at 0 100
\tarrow from 2 77.52 to -2 77.52
\tarrow from -2 122.48 to 2 122.48
}{
\bsmbl\plotsymbolspacing=7\tdim
\startrotation by 0 1
\plot -7 102.03 7 102.03 /
\stoprotation
\linethickness=0pt
\putrule from 0 -115 to 0 115
\putrule from -105 0 to 105 0}
\endpicture$$
\caption{\figsize\sf\label{fig-2}The leading diagram 
contributing to the vacuum energy.}%
\end{figure}}

The resulting vacuum energy and scalar masses are then given by
\begin{align}
  \label{eq:4}
  \Lambda &\sim \sin^2 \theta
  \left(\frac{g_1}{4\pi}\right)^2\dots\left(\frac{g_N}{4\pi}\right)^2
  \frac{M^4}{16\pi^2} \\
  \label{eq:4a}
  m_\phi^2 &\sim \sin^2 \theta
  \left(\frac{g_1}{4\pi}\right)^2\dots\left(\frac{g_N}{4\pi}\right)^2
  M^2
\end{align}
Since these are UV effects the couplings are to be evaluated at the
scale $M$.  The sign and precise values of these quantities depend on
the details of the UV physics; here we have indicated the natural size
assuming the cut-off procedure preserves local supersymmetry in theory
space.

The breaking of supersymmetry through $\arg J$ is clearly explicit:
there are no conserved supercharges and there is no goldstino.
Nevertheless as long as supersymmetry is preserved locally in theory
space, broken only by a global ``twist'' in the couplings,
supersymmetry breaking is controllably small, parametrically
exponentially small in $N$.

{\figsize%
\begin{figure}[htb]
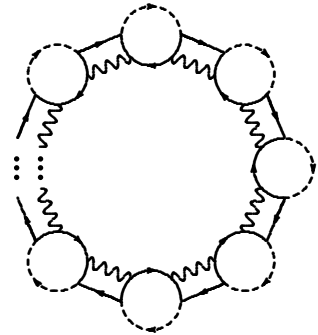

$$\beginpicture
\setcoordinatesystem units <1\tdim,1\tdim>
\stpltsmbl
\moose{\tarrow from -17.5 102.03 to 2 102.03
\setlinear
\plot -17.5 102.03 17.5 102.03 /
\setquadratic
\plot -17.5 83.79 -15 79.79 -12.5 83.79 
-10 87.79 -7.5 83.79 -5 79.79 
-2.5 83.79 0 87.79 2.5 83.79 5 79.79 7.5 83.79 
10 87.79 12.5 83.79 15 79.79 17.5 83.79
/
}{
\setdashes <4\tdim>
\circulararc 180 degrees from 22.48 100 center at 0 100
\setsolid
\circulararc 180 degrees from -22.48 100 center at 0 100
\tarrow from 2 77.52 to -2 77.52
\tarrow from -2 122.48 to 2 122.48
}{
\bsmbl
\plotsymbolspacing=7\tdim
\startrotation by 0 1
\setlinear
\plot -7 102.03 7 102.03 /
\plot -7 83.79 7 83.79 /
\stoprotation
\linethickness=0pt
\putrule from 0 -115 to 0 115
\putrule from -105 0 to 105 0
}
\endpicture$$
\caption{\figsize\sf\label{fig-3}The same as fig. 2 for a non-Abelian
theory.}
\end{figure}}

The same phenomena occurs for the $SU(k)^N$ model. In this case, the
leading supersymmetry breaking diagrams are shown in fig. 3, and the
vacuum energy and scalar masses are given by
\begin{align}
  \label{eq:5}
  \Lambda &\sim \sin^2 \theta
  \left(\frac{g_1}{4\pi}\right)^4\dots\left(\frac{g_N}{4\pi}\right)^4
  \frac{M^4}{16\pi^2} \\
  m_\phi^2 &\sim \sin^2 \theta
  \label{eq:5a}
  \left(\frac{g_1}{4\pi}\right)^4\dots\left(\frac{g_N}{4\pi}\right)^4
  M^2
\end{align}

Note that if we send $N\to\infty$ while holding $M$ fixed,
supersymmetry is restored. For the non-abelian case we may contemplate
sending $M\to\infty$, holding the $SU(k)_i$ scales $\Lambda_i$ fixed.
We can take the $M\to\infty, N\to\infty$ limit such that $m_\phi^2
\sim \left(g^{(1)}\dots g^{(N)}\right)^4 M^2$ remains fixed. In this
case supersymmetry violating effects appear only through the scalar
masses $m_\phi$, just as in a theory with only soft supersymmetry
breaking masses added to the Lagrangian.  However in this theory an
exponentially large hierarchy between these these supersymmetry
breaking masses and the cut-off is naturally obtained.  For instance,
$m_\phi/M \sim 10^{-16}$ for $N\sim 8$ and $g,\theta\sim 1$.

The condition $\vert h_i\vert, \vert h'_i\vert = g_i$ was assumed for
the bare couplings.  At energies beneath the cut-off, these couplings
run, and will no longer be equal even in magnitude. The differences,
like all supersymmetry breaking effects, will be proportional to $J$,
and vanish exponentially for large $N$.  How then does a low energy
observer distinguish supersymmetry breaking via a global twist in
theory space from small explicit ``local'' supersymmetry breaking?
The observer would simply run the couplings to higher energies and
notice that there is {\em some} scale at which the magnitude of all $2
N$ Yukawa couplings are equal to the corresponding gauge couplings.

\section{Relation to Scherk-Schwartz}

So far we have discussed the consequences of twisted supersymmetry
near the origin of the classical moduli space.  It is interesting to
explore what happens around a generic point in the classical moduli
space, which may include regions where the theory space turns into
extra dimensions. In particular, since our supersymmetry breaking is
associated with a phase around the circle, we expect that twisting
reduces to Scherk-Schwartz supersymmetry breaking in those regions of
classical moduli space where extra dimensions are generated.  Consider the
$U(1)^N$ model, and expand the theory around the classical vacuum
where $\Phi_1 = \dots = \Phi_N = \phi$. In the supersymmetric limit,
this generates a fifth dimension. In the presence of a non-trivial
phase $\theta$, the spectrum of the theory already exhibits bose-fermi
splitting at tree-level. Since the phase only appears in the gaugino
Yukawa couplings, the bosonic spectrum is unaffected. But the fermion
mass terms $\lambda M_F \psi$ will have phases. By making phase
redefinitions, we can always choose $h_i^*,h'_i = g_i e^{i \theta/N}$.
Then the mass squared matrix for the fermions $(M^2_F) \equiv
M_F^\dagger M_F$ becomes
\begin{equation}
(M^2_F)_{ij} = \frac{1}{a^2} \left(2 \delta_{i,j} -e^{-2 i \theta/N}
\delta_{i,j+1} - e^{2 i \theta/N} \delta_{i,j-1}\right) 
\end{equation}
The spectrum of bosons and fermions in the theory is then 
\begin{align}
  \label{eq:6}
  m_{B,n}^2 &= \left(\frac{2}{a}\right)^2
  \sin^2\left(\frac{na}{2R}\right) \\ 
  \label{eq:6a}
  m_{F,n}^2 &= \left(\frac{2}{a}\right)^2 \sin^2\left(\frac{na}{2R} +
    \theta \right)
\end{align}
For the modes lighter than $1/a$, this is precisely the spectrum from
Scherk-Schwartz supersymmetry breaking in a fifth dimension where the
fermions pick up a phase $e^{2 i \theta}$ around the circle.

How is this tree-level non-supersymmetric bose-fermi splitting
consistent with our previous analysis, which argued that all
supersymmetry breaking effects must involve $J$? To clarify this we
examine the bose-fermi mass splitting for the lightest mode in the
small $\theta$ limit:
\begin{equation}
  \label{eq:1}
  m_{F,0}^2 - m_{B,0}^2 \simeq  \frac{4}{a^2} \theta^2 =  4 g^2
  \phi^2 \theta^2
\end{equation}
This does not appear to vanish like $g^{2N}$ as our previous arguments
would seem to suggest. But consider the case of general gauge couplings $g_i$,
not all equal. In this case the mass splitting is
\begin{equation}
  \label{eq:8}
  m_{F,0}^2 - m_{B,0}^2 \simeq \frac{4 N (g_1^2 \phi^2)\cdots(g_N^2
    \phi^2)  \theta^2}{\sum_{k=1}^N (g_1^2
    \phi^2)\cdots\widehat{(g_k^2\phi^2)}\cdots(g_N^2\phi^2)}
\end{equation}
where the hatted term is to be excluded from the product.  Here the
numerator is explicitly proportional to $(\text{Re} J - g_1^2\dots
g_N^2)$ just as are the UV sensitive contributions of the previous
section, and supersymmetry breaking still vanishes if any one gauge
coupling is zero.  However in this case there are new {\em infrared}
mass scales, set by $g_i \phi$ which appear in the denominator, and
enhance the effect. This phenomenon is well-known and ubiquitous.  For
example in the Standard Model an invariant measure of CP violation,
proportional to many small Yukawa couplings, is of order $10^{-20}$.
This is enhanced by small infrared mass differences to produce the
observed CP violation in the Kaon system of order $10^{-3}$.

In the $SU(k)^N$ example, along the directions of classical moduli space where
$\Phi_i = \phi_i \mathbf{1}_k$, tree-level bose-fermi mass splittings
may  also be produced. Here the fermi-bose mass splitting for the
lightest mode in the small $\theta$ limit is
\begin{equation}
  m_{F,0}^2 - m_{B,0}^2 \simeq \frac{4 N (g_1^2 \phi_1^2)\cdots(g_N^2
    \phi_N^2) \theta^2}{\sum_{k=1}^N (g_1^2
    \phi_1^2)\cdots\widehat{(g_k^2\phi_k^2)}\cdots(g_N^2\phi_N^2)} 
\end{equation}
Note that if any one of the $\phi_i$ vanishes, bose-fermi degeneracy
is restored at tree level. This highlights the global nature of the
supersymmetry breaking---only in regions of classical moduli space where all
links are non-zero does tree-level supersymmetry breaking appear.

Since supersymmetry is broken, we do not expect the classical moduli
space to remain flat; quantum corrections will generate a potential
for the moduli. For the $U(1)^N$ example the UV contributions of the
previous section generate a quadratic potential $m_\phi^2 \phi^2$ with
$m_\phi^2 \sim g^{2N} M^2$ On the other hand, away from the origin of
classical moduli space there are tree-level bose-fermi splittings
proportional to $\phi$, and so a Coleman-Weinberg potential for $\phi$
is generated already at 1-loop.  Since all UV sensitive contributions
appear first at $N$ loops, the 1-loop Coleman-Weinberg potential
cannot contain UV divergences for $N>2$, with at most log divergences
for $N=2$.  This can be verified by computing the 1-loop potential
explicitly. The result is indeed cut-off independent for $N>2$ and is
given by \cite{Arkani-Hamed:2001nc}:
\begin{equation}
  \label{eq:7}
  V_{\text{1-loop}}(\phi) = -\frac{3 g^4 \phi^4}{4 \pi^2}
  \sum_{j=1}^{\infty} \frac{\sin^2(n N \theta)}{n(n^2 N^2 - 1)(n^2 N^2
    - 4)}
\end{equation}
This corresponds to the $1/R^4$ potential that is generated for the
radius modulus in Scherk-Schwartz models.  If $m_\phi^2$ is negative,
$\phi$ rolls off to the cut-off $M$, while if the sign is positive, the origin
is meta-stable, with scalar masses hierarchically smaller than $M$.

The $SU(k)^N$ model is similar. If all the $\phi_i$ are non-zero a
1-loop Coleman-Weinberg potential is generated. If any one of the
$\phi_i$ are zero, there is no tree-level bose-fermi splitting and the
1-loop potential vanishes. In this case the leading infrared contribution to
the potential first appears at 3 loops. In general if $r$ of the
$\phi_i$ are zero this potential first appears at $2r+1$ loops.

The negative Coleman-Weinberg potential in the $U(1)^N$ model arose because
only the fermionic spectrum was modified by $\theta$.  The addition of
other degrees of freedom, such as ``hopping'' chiral superfields, can
affect both bosons and fermions. In this case the Coleman-Weinberg
potential can be positive.  We may also construct models in which the
$\phi$ flat direction is lifted at tree level by additional
interactions.  These theories may then have stable minima with
$g \phi \sim g^N M$, hierarchically below $M$. This dynamically
generates an extra dimension with Scherk-Schwartz supersymmetry
breaking, but with a stabilized radius hierarchically larger than
$M^{-1}$.

Scherk-Schwartz supersymmetry breaking via boundary conditions around
a fifth dimension, although explicit, is usually said to be very soft at
high energies, and therefore insensitive to the physics at the cut-off
of the five-dimensional theory
\cite{Antoniadis:1998zg,Barbieri:2000vh,Arkani-Hamed:2001mi}.  It is
most convenient to use position space for the fifth dimension and
momentum space for the remaining dimensions. Just as finite
temperature boundary conditions introduce an exponential Boltzmann
factor, compactification on a circle of radius $R$ with
Scherk-Schwartz boundary conditions introduces factors
$\exp(-\abs{p_4} R)$, in (Euclideanized) Feynman diagrams involving
supersymmetry breaking cutting off loop momenta at $1/R$.  This makes
supersymmetry breaking radiative corrections finite and calculable, up
to effects from higher-dimension local operators power-suppressed by
the cut-off.  Divergences are associated with the {\em short distance}
structure of the theory and, provided that the physics at the cut-off
$\Lambda$ is local on the scale $\Lambda^{-1}$, the breaking of
supersymmetry by boundary conditions can not introduce new
divergences.

In momentum space for the fifth dimension the essential physics of
locality is obscured.  Since the higher-dimensional physics becomes
strongly coupled, not all the Kaluza-Klein modes can be reliably
included in computing radiative corrections, but  sharply cutting off the
KK sum at some maximum mode gives UV-divergent supersymmetry breaking
effects even at low loop order.  However, sharp momentum cut-offs also
give rise to power-suppressed (rather than exponentially suppressed)
non-local effects in position space.  If any fully local cut-off
procedure is used, these divergent effects disappear.

The situation is especially clear in our concrete UV completions of
these models.  In order to mimic a continuum field-theoretic
dimension, we take the large $N$ limit. For any finite $N$, the KK
tower is finite, containing only $N$ modes, and there are no UV
divergent supersymmetry breaking effects at low loop order.  In the
KK sum this follows from  cancellations associated with the
specific $\sin^2 (n a/2R), \sin^2(na/2R + \theta)$ spectrum. 
But the momentum space KK calculation is a cumbersome way to deduce what
is obvious in theory space (transmuted here into an extra dimension),
guaranteed by  ``local'' supersymmetry.

There is a more interesting puzzle here. While there are no UV
divergent effects at low-loop order, there {\em are} power-UV
divergent effects at $N$ loops. What is the interpretation of these
effects from the low-energy, five dimensional point of view?  In
taking the large $N$ limit we wish to keep the effective five
dimensional gauge coupling $1/g_5^2$ and the compactification radius
$R$ fixed. Since the five dimensional gauge coupling is $1/g_5^2 =
1/(a g^2)$, this keeps the dimensionless ratio $g_5^2/R = g^2/N \equiv
g^2_{LE}$ fixed, and the infinite $N$ limit pushes the theory into
strong coupling where our picture of the physics breaks down.
Nevertheless we can take $ g \sim 4 \pi$ corresponding to $N
\alpha_{LE}/4\pi \sim 1$.  The power UV divergent effects for
{\it e.g.} scalar masses are
\begin{equation}
  m_\phi^2 \sim \left(\frac{\alpha}{4\pi}\right)^N M^2 \sim
  \left(\frac{\alpha_{LE} N}{4\pi} \right)^N M^2  
\end{equation}
As $N$ is increased holding $\alpha_{LE}$ fixed, this function
decreases till $\alpha_{LE} N \sim 1$, beyond which it increases
again; but beyond this point we don't trust our picture of the physics
anyway. So, taking $N$ as large as we can to best mimic the continuum
extra dimension, the size of the power-supersymmetry violating effects is of
order
\begin{equation}
  m_\phi^2 \sim e^{-4\pi/\alpha_{LE}} M^2 \sim e^{- 2\pi R \Lambda_{5}} M^2
\end{equation}
where $\Lambda_{5} \sim 16 \pi^2/g_5^2$ is the naive scale at which the
low-energy five dimensional theory becomes strongly coupled. 

In matching to the five dimensional low energy theory, there is a
non-local contribution to the scalar mass, deriving from the non-local
breaking of supersymmetry in the UV theory.  As befits a non-local
effect, it is exponentially small $\sim e^{-R \Lambda_{5}}$.  In the
usual effective field theory with extra dimensions, it is assumed that
the five dimensional cut-off $\Lambda_5$ sets {\em all} the
short-distance scales in the theory, so the coefficient of such a
non-local effect would be $\Lambda_5^2$ and it's size would be
miniscule. There are many UV completions where such an expectation is
justified, and in our UV completion the same conclusion is reached if
we choose $M \sim \Lambda_5$.

However, in our UV completion, just as the extra dimension itself has
no fundamental significance, neither does the na\"\i{}ve five dimensional
cut-off. The true UV cut-off $M$ can be much higher than
$\Lambda_5$. In the limit of exact supersymmetry, the presence of the
higher cutoff $M$ would be invisible to the low-energy five dimensional
observer.  But, in the presence of the explicit supersymmetry breaking
associated with the compactification of the extra dimension, the much
higher energy scales that were hidden due to supersymmetric
cancellations  become relevant, and the size of the supersymmetric
breaking is much larger than the low-energy five dimensional observer
would expect to be associated with a large-distance effect in the
fifth dimension.

This violation of low-energy expectations for the size of
supersymmetry breaking is intriguing, and provides a toy example of part of a
scenario advocated by Banks to address the cosmological constant
problem \cite{Banks:2000fe}. In Banks' picture all of supersymmetry
breaking is cosmological in origin, associated with a deSitter space
with a curvature of order the present Hubble radius.  The
phenomenological problem is then to obtain supersymmetry breaking mass
splittings much larger than the na\"\i{}ive classical expectation
$\sim 10^{-3}\text{eV}$.  Radiative corrections seem unable to produce
such large splittings, being only logarithmically sensitive to the
Planck scale.  Banks conjectures that, because of now inexact
supersymmetric cancellations, very high energy states above the Planck
scale contribute and dramatically modify the size of the scalar
masses.  As we have seen, exactly this physics occurs in our models,
if the ``breaking of supersymmetry from deSitter space'' is replaced
by ``breaking of supersymmetry by boundary conditions in a fifth
dimension'', and if the Planck scale is replaced by the naive five
dimensional cutoff $\Lambda_5$. A related toy example of this sort was
presented in \cite{Banks:2000gw}.

\section{``Local'' supersymmetry and Homology}

In our previous models, the requirement of ``local'' supersymmetry
was simply satisfied, and the supersymmetry breaking rephase invariant
was easy to identify.  We now wish to extend our discussion to a
general theory characterized by a theory space.  As before, we
consider assigning arbitrary phases in the gauge-Yukawa couplings
involving the link fields.

It will prove convenient to define a {\em directed} link $l =
(ij)_\alpha$ with a line pointing from $i$ to $j$.  The gauge-Yukawa
couplings for the components of the chiral superfields $\Phi_l$
associated with the link $(i,j)_\alpha$ are
\begin{equation}
  \sqrt{2} \phi^{\dagger}_l h_l \lambda_i \psi_l + 
  \sqrt{2} \phi^{\dagger}_l h'_l \lambda_j \psi_l
\end{equation}
Once again, we will demand that the magnitudes $\abs{h_l} = g_i$
and $\abs{h'_l} = g_j$, but allow these couplings to have phases.
Under field rephasings the product $(h_l^* h'_l)$ is only
affected by the gaugino rephasings $\lambda_i \to \omega_i \lambda_i$,
transforming as
\begin{equation}
  (h_l^* h'_l) \to \omega_i^* (h_l^* h'_l) \omega_j
\end{equation}
This motivates the  association of a phase $U_l$ with each $l$, defined as 
\begin{equation}
  h_l^* h'_l \equiv g_i g_j U_l 
\end{equation}
Under the gaugino rephasings, $U_l$ transforms as
\begin{equation}
  U_l \to \omega^*_i U_l \omega_j
\end{equation}
Note that $U_l$ transforms just as a link variable for a $U(1)_R$
lattice gauge theory, under which the gauginos on the sites are
charged.

Suppose that by gaugino rephasings we could set all $U_l = 1$.  Then,
$h_l,h'_l$ have the same phase for every link, and we could remove
this phase by rephasing the components of $\Phi_l$, without affecting
any of the other couplings in the theory. Hence any breaking of
supersymmetry must be associated with non-trivial rephase invariants
constructed from the $U_l$.  From the lattice gauge theory analogy, we
expect these invariants to be associated with closed Wilson lines. We
can characterize the rephase invariants precisely by using
elementary notions from topology.  Let $(l_1,l_2, \dots, l_{N_l})$ be
a list of all the directed links in theory space.  The most general
candidate for an invariant phase is a product of $U_l$s of the
form $U_{l_1}^{a_1} U_{l_2}^{a_2} \cdots U_{l_{N_l}}^{a_{N_l}}$, where
the $a$s are integers.  Define a ``chain'' $c$ to be a formal sum
of the form $c= a_1 l_1 + a_2 l_2 + \dots a_N l_N$, and define a
function on chains
\begin{equation}
  V(c) = U_{l_1}^{a_1} U_{l_2}^{a_2} \cdots U_{l_{N_l}}^{a_{N_l}}
\end{equation}
It follows that for two chains $c_1,c_2$, $V(c_1) V(c_2) = V(c_1 +
c_2)$.   One trivial set of $c$s, for
which $V(c)=1$, are those traversing a path both forwards and backwards,
{\it e.g.} $c=(ij)_\alpha + (ji)_\alpha$. We will therefore define
$-(ij)_\alpha = (ji)_\alpha$ so that all such chains vanish. In order
to characterize the non-trivial rephase invariants, it is helpful to
define a linear ``boundary'' operator $\partial$ which acts on a
directed link field $l = (ij)_\alpha$ as $\partial \, (ij)_\alpha =
(j) - (i)$, where $(j) - (i)$ is also understood as a formal sum.  It is
then easy to see that $V(c)$ is rephase invariant if and only if $c$
is a ``closed'' chain, satisfying $\partial c = 0$. The associated
rephase invariant combination of couplings whose phase is $V(c)$ is
\begin{equation}
  J(c) = \tilde{g}_{l_1}^{2 a_1} \cdots \tilde{g}_{l_{N_l}}^{2
    a_{N_l}} V(c)
\end{equation}
where for each $l = (ij)_\alpha$ we define $\tilde{g}^2_l \equiv g_i g_j$.

Exact supersymmetry requires $V(c) = 1$ for all closed chains $c$.
But we wish to impose the weaker condition of ``local'' supersymmetry.
To do this, we need to make a choice for some set of sites and links
that are ``locally connected''.  Such a choice can be specified by a
non-zero closed chain of the form $(i_1 i_2)_{\alpha_1} + (i_2
i_3)_{\alpha_2} + \dots + (i_{N_p} i_1)_{\alpha_{N_p}}$ which
includes the ``locally connected'' sites and links. We define a
plaquette $p$ to be an object $p = (i_1 i_2)_{\alpha_1} (i_2
i_3)_{\alpha_2} (i_{N_p} i_1)_{\alpha_{N_p}}$, whose boundary is
$\partial p = (i_1 i_2)_{\alpha_1} + (i_2 i_3)_{\alpha_2} + \cdots +
(i_{N_p} i_1)_{\alpha_{N_p}}$.  Just as we defined a chain $c$ to be a
formal sum of directed links, we define a surface $s$ to be a formal
sum of plaquettes.

We can now state the condition of ``local'' supersymmetry precisely:
we require that the phases from any given ``locally connected'' set of
sites and links characterized by $p$ can be removed by rephasing the
fields. If $V(\partial p) \neq 1$, this is impossible. So we require
that for all plaquettes $p$, $V(\partial p) = 1$.  Since $V(c_1 + c_2)
= V(c_1) V(c_2)$, this immediately implies that for any surface $s$,
$V(\partial s) = 1$.  Therefore, if we have two closed chains
$c_1,c_2$ such that $c_1 - c_2 = \partial s$ for some surface $s$,
$V(c_1) = V(c_2)$. This is also familiar from the lattice analogy. The
requirement of ``local'' supersymmetry in this case is the absence of
$U(1)_R$ flux through the plaquettes, and therefore if two closed
paths form the boundary of a region consisting of plaquettes, the
associated Wilson loops will be equal.

This motivates the introduction of an equivalence relation between
closed chains where $c_1,c_2$ are ``homologous'' $c_1 \sim c_2$ if
$c_1 - c_2 = \partial s$ for some $s$. Then for every equivalence
class $[c]$ of homologous chains there is an associated phase $V([c])$
common to all the chains in $[c]$.  These equivalence classes form an
additive group with the addition rule $[c_1] + [c_2] = [c_1 + c_2]$,
the first Homology group $H_1(T,P)$ associated with the theory space
$T$ and choice of plaquettes $P$, and the phases $V([c])$ must be a
representation of $H_1(T,P)$.  If $H_1(T,P)$ is trivial, then the
requirement of ``local'' supersymmetry is enough to ensure exact
supersymmetry; if not we can introduce supersymmetry breaking in a way
consistent with ``local'' supersymmetry.

Let us look at a few examples, beginning with our simple circular
theory spaces. All closed chains are of the form $n c_1$ where $c_1 =
(12) + (23) + \cdots (N1)$. The only possible plaquette is $p=(12)(23)
\dots (N1)$ with boundary $\partial p = c_1$. The possible choices
for $P$ are then either the empty set or $p$. Clearly the latter does
not correspond to any reasonable notion of locality, since with this
definition all sites are locally connected even as $N \to \infty$. We
therefore choose $P$ to be empty, which is another way of saying that
our requirement of ``local'' supersymmetry is automatic in this
theory. Since all closed chains are multiples of $c_1$ and there are
no plaquettes, $H_1(T,P)=Z$ with elements $[n c_1]$, and the phases
$V([n c_1]) = e^{i n 2 \theta}$.

{\figsize%
\begin{figure}[t]
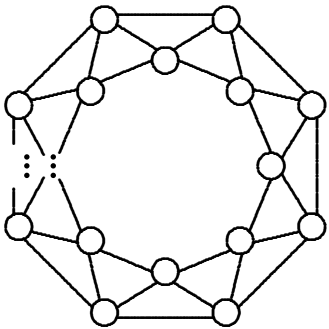

$$\beginpicture
\setcoordinatesystem units <1\tdim,1\tdim>
\stpltsmbl
\moose{\circulararc 360 degrees from 10 120 center at 0 120
\plot -21.4 77.7 21.4 77.7 /
}{\circulararc 360 degrees from 10 80 center at 0 80
\plot -36.7 114.7 36.7 114.7 /
\plot -36.7 107 -6 88 /
\plot 36.7 107 6 88 /
}{\startrotation by 0 1
\plot -36.7 114.7 -16.7 114.7 /
\plot 36.7 114.7 16.7 114.7 /
\plot -36.7 107 -9 92 /
\plot 36.7 107 9 92 /
\bsmbl\plotsymbolspacing=7\tdim
\plot -7 105 7 105 /
\plot -7 85 7 85 /
\stoprotation
\linethickness=0pt
\putrule from 0 -115 to 0 115
\putrule from -105 0 to 105 0}
\endpicture$$
\caption{\figsize\sf\label{fig-4}Ribbon theory space}\end{figure}}

Now consider the ``ribbon'' theory space shown in fig. 4.  There are
many possible choices for the plaquettes in $P$; the most natural one
is the triangles connecting nearest neighbor sites with clockwise
orientation, by which we mean $p=l_1 l_2 l_3$ where the links
$l_1,l_2,l_3$ wind clockwise around a triangle in the figure.  With
this choice it is clear that all closed chains are homologous to
multiples of {\it e.g.} the chain going around the inner polygon, and
again $H_1(T,P) = Z$.

Next, consider the theory space of fig. 5; the presence of the group
in the center makes the theory ``non-local'' in theory space, in the
sense that every point is connected to every other point through at
most two links.  If we choose the plaquettes to contain the triangles
as before, the theory space is topologically a disk and all chains are
homologous to zero, so no supersymmetry breaking phases are allowed.
If we chose $P$ to be empty instead, in addition to the non-trivial
closed chains winding around the circle, there are new ``small''
chains winding around the individual triangles, and the supersymmetry
breaking rephase invariants are unsuppressed as we take $N \to
\infty$.  Note that the rephase invariants associated with the small
cycles are all proportional to the gauge coupling of the center group,
so that these supersymmetry breaking invariants vanish as this gauge
coupling vanishes.  In this limit
the center point is effectively removed from the theory space and
we revert to the circular model.  

{\figsize%
\begin{figure}[htb]
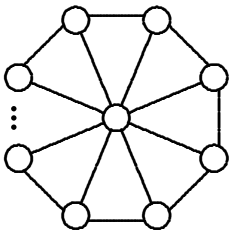

$$\beginpicture
\setcoordinatesystem units <1\tdim,1\tdim>
\stpltsmbl
\moose{\circulararc 360 degrees from 10 80 center at 0 80
\plot 0 10 0 70 /
}{
\plot -21.38 77.7 21.38 77.7 /
}{\circulararc 360 degrees from 10 0 center at 0 0
\bsmbl\plotsymbolspacing=7\tdim
\startrotation by 0 1
\plot -7 77.7 7 77.7 /
\stoprotation
\linethickness=0pt
\putrule from 0 -100 to 0 100
\putrule from -105 0 to 105 0}
\endpicture$$
\caption{\figsize\sf\label{fig-5}A non-local theory space.}
\end{figure}}

Finally, consider the theory space of fig. 6, consisting of a site
at the center with $2 N$ lines radiating outward, intersecting $k$
concentric circles each with $2N$ sites.  
{\figsize%
\begin{figure}[htb]
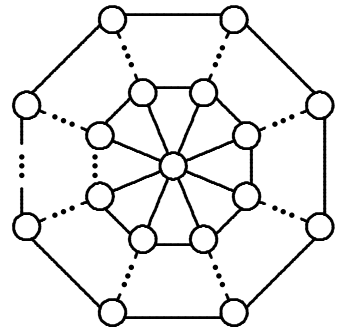

$$\beginpicture
\setcoordinatesystem units <1\tdim,1\tdim>
\moose{\stpltsmbl\plotsymbolspacing=.6\tdim
\circulararc 360 degrees from 10 120 center at 0 120
\circulararc 360 degrees from 10 60 center at 0 60
\plot 0 110 0 105 /
\plot 0 70 0 75 /
\plot 0 10 0 50 /
\bsmbl\plotsymbolspacing=7\tdim
\plot 0 83 0 97 /
}{\stpltsmbl\plotsymbolspacing=.6\tdim
\plot -36.7 114.7 36.7 114.7 /
\plot -13.7 59.26 13.7 59.26 /
}{\stpltsmbl\plotsymbolspacing=.6\tdim
\circulararc 360 degrees from 10 0 center at 0 0
\startrotation by 0 1
\plot -36.7 114.7 -16.7 114.7 /
\plot 36.7 114.7 16.7 114.7 /
\bsmbl\plotsymbolspacing=7\tdim
\plot -7 114.7 7 114.7 /
\plot -7 59.26 7 59.26 /
\stoprotation
\linethickness=0pt
\putrule from 0 -115 to 0 115
\putrule from -105 0 to 105 0}
\endpicture$$
\caption{\figsize\sf\label{fig-6} Spider web theory space.}
\end{figure}}
The most natural choice of plaquettes consists of all clockwise
oriented triangles and squares in the figure. As it stands, this space
is also topologically a disk with a trivial $H_1(T,P)$.  However, we
can make the model more interesting by identifying diametrically
opposite sites (and the links connecting them) on the boundary, {\it
  i.e.}  identifying the site $i$ with $i+N$ on the $k$th circle.  The
resulting space is topologically the same as the projective plane
$RP^2$, with a first homology group $H_1(RP^2)=Z_2$.  We can see this
directly as follows. All closed chains are clearly homologous to
multiples of the chain $c_1$ that traverses the circumference of the
largest circle, starting from a given site $i$ to $i+N$. Consider the
boundary $\partial s$ of the surface $s$ which is the sum of all the
plaquettes in the theory. The contribution to $\partial s$ from all
the links on the interior of the diagram cancel, and we are left with
the sum of the links on the outside, which is equal to $2 c_1$. That
is, $2 c_1 = \partial s$, and so $[2 c_1] = [0]$.  Therefore $H_1(T,P)
= Z_2$, and the only possible supersymmetry breaking rephase invariant
is $V([c_1]) =-1$.  This is pleasing: in our examples with a circle,
the supersymmetry violating phase was arbitrary. Here, the requirement
of ``local'' supersymmetry either demands exact supersymmetry or a
maximal supersymmetry violating phase.  This construction clearly
generalizes: if we take $m N$ sites on the boundary circle and again
identify the sites $i$ with $i+N$, the homology group is $Z_m$, and
the rephase invariant phases are the $m$th roots of unity.

\section{Twisting ${\cal N}=2$ Supersymmetry}

Supersymmetry may be twisted in ${\cal N}=2$ models as well.
Including phases in the gauge-Yukawa couplings as before, the ${\cal
  N}=2$ supersymmetry would break all the way down to ${\cal N}=0$.
But we can also 
construct models which leave an ${\cal N}=1$ supersymmetry intact.
Consider the model of figure 1 as an ${\cal N}=2$ supersymmetric
$U(1)^N$ field theory. In this case each link corresponds to a
hypermultiplet which, in ${\cal N}=1$ language, contains two chiral
multiplets $H^c, H$, while the vector multiplet at each site contains
a chiral adjoint superfield $X$. The superpotential is
\begin{equation}
  W=  \sum_{i=1}^N   \sqrt{2} H^c_i \left( f_i X_i -
    f'_{i-1} X_{i-1}\right) H_i
\end{equation}
${\cal N}=2$ supersymmetry requires that these Yukawa couplings are
equal to the gauge couplings: $f_i,f'_i=g_i$.  Just as in our ${\cal
  N}=1$ example, we allow for arbitrary phases in these couplings. By
the same arguments used in the ${\cal N}=1$ example, we see that the
combination of couplings analogous to $J$ is the unique rephase
invariant measure of supersymmetry breaking in these couplings, in
this case from ${\cal N}=2$ to ${\cal N}=1$. The exact ${\cal N}=1$
supersymmetry eliminates any power sensitivity to the UV, but the
${\cal N}=2$ non-renormalization theorems will be violated at $N$-loop
order. For example ${\cal N}=2$ supersymmetry forces the
hypermultiplet wave-function renormalization to vanish. Here, the
wave-function renormalization is non-zero, but small:
\begin{equation}
  \label{eq:9}
  \log Z_H \sim \left(\frac{\alpha}{4\pi}\right)^N
\end{equation}
Note that the generalization to ${\cal N}=2$ $SU(k)^N$ theory is a
conformal field theory before twisting. Upon twisting the conformal
symmetry is broken, but only through wave-function renormalization
which appears at $N$ loops! This provides a natural example of a
``walking'' gauge theory, with gauge couplings evolving arbitrarily
slowly beneath the cut-off.

We can also break ${\cal N}=2$ to ${\cal N}=0$ in a different way.
Imagine removing the chiral multiplet $X_i$ from the $i$th site on the
chain.  This would break the supersymmetry down to ${\cal N}=1$.
Alternatively we could remove the complex scalar component of $X_i$
and the gaugino component $\lambda_i$ of the vector multiplet on this
site.  This still preserves an ${\cal N}=1$ supersymmetry, where the
fermionic component of $X_i$ takes the place of the gaugino.  But now
consider a finite chain of $N$ gauge groups rather than a circle. If
we remove the chiral multiplet $X_N$ from the rightmost site {\em as
  well as} the complex scalar from $X_1$ and the gaugino $\lambda_1$
from the leftmost site we completely break supersymmetry.  Locally we
preserve an ${\cal N}=2$ supersymmetry, where the two end sites each
preserve a different ${\cal N}=1$ subgroup of the ${\cal N}=2$
supersymmetry. Consequently fully non-supersymmetric effects must
involve all couplings from each site, again appearing only at $N$
loops.

\section{Conclusions and Outlook}

The exploration of theory space has revealed surprising new phenomena.
In special cases, field-theoretic extra dimensions with their
familiar dynamics arise from theory space at low energies. More
generally there are new features that are intrinsic to theory space
itself.  Supersymmetric field theories provide a particularly clear
illustration of this point. Only in some regions of classical moduli
space does theory space transmute into extra dimensions, but the
underlying theory space plays a deeper role.  The structure of sites
and links together with a notion of ``locality'' endows theory space
with a topological significance irrespective of the choice of vacuum.
If this topology is non-trivial it is possible to preserve
supersymmetry locally in theory space, but break it through
topological obstructions.  This breaking manifests itself differently
in different regions of the classical moduli space. At the origin
supersymmetry breaking effects appear only at $N$-loop order, while in
regions of classical moduli space where the entire theory space
reduces to extra dimensions the breaking appears at tree-level,
reducing to the Scherk-Schwartz mechanism of supersymmetry breaking.
In general regions of classical moduli space only certain areas of the
theory space reduce to extra dimensions, and supersymmetry breaking
effects involve perturbative loop factors for all links outside these
areas, interpolating between the unsuppressed breaking of extra
dimensions and the exponentially small breaking of bare theory space.

Even when  twisted supersymmetry reduces to
Scherk-Schwartz breaking, theory space gives new insight.  The models
presented here provide a UV completion \cite{Arkani-Hamed:2001ca} of
the Scherk-Schwartz mechanism, allowing investigation of the physics
at energies near and above the na\"\i{}ve cut-off of the
higher-dimensional theory.

Twisted supersymmetry breaking has obvious phenomenological
applications.  When theory space reduces to extra dimensions, twisted
supersymmetry reduces to standard Scherk-Schwartz supersymmetry
breaking with the potential added benefit of naturally generating a
large hierarchy between the compactification scale $1/R$ and the
cut-off $M$.  But non-extra-dimensional examples are also possible.
We may imagine a theory space in which the
non-trivial twist resides in some portion while the Standard Model
fields reside in another.  Supersymmetry breaking then appears to be
communicated to the Standard Model from a ``hidden sector''. Since
twisted supersymmetry breaking effects are suppressed by many loop
factors, this can also generate a large hierarchy.

We have so far ignored gravity.  If we na\"\i{}vely add
four-dimensional supergravity, these interactions act non-locally in
theory space and therefore supersymmetry breaking need not involve the
large number of gauge couplings present in our rephase invariants.
For example there are direct gravitational couplings of twisted sector
supersymmetry breaking to Standard Model fields which gives Standard
Model scalar masses of size
\begin{equation}
  \label{eq:10}
  m^2 \sim \frac{M^6}{M_{\text{Planck}}^4}
  \left(\frac{\alpha}{4\pi}\right)^2  
\end{equation}
For the UV scale $M$  smaller than $M_{\text{Planck}}$ these
effects can be quite small,  even for $M$ far above the TeV scale.
For example for $\alpha/(4\pi) \sim 10^{-2}$, these masses
are smaller than 1 TeV for all $M < 10^{11}$ TeV. 
Gravitational interactions will also generate  supersymmetry breaking
scalar masses  directly in the twisted sector, of size
\begin{equation}
  \label{eq:11}
  m_\phi^2 \sim \frac{M^4}{M_{\text{Planck}}^2} \left(\frac{\alpha}{4
      \pi}\right)^2 
\end{equation}
These in turn will generate supersymmetry breaking in the Standard
Model sector through the links in theory space that connect the
Standard Model to the twisted sector. Again, even for a small number
of links and a moderate $N$ these effects are smaller than the direct
effects from twisted supersymmetry.

In previous work we used theory space to accelerate the supersymmetric
unification of gauge couplings to much lower energies.  It is
therefore natural to try and combine accelerated unification with
twisted supersymmetry. There is also a phenomenological hint to
do so.  On the one hand, gauge coupling unification suggests  a high
fundamental scale. On the other, if supersymmetry breaking is present
at such a high scale, radiative corrections to the Higgs mass from
top squarks are enhanced by a large logarithm, so that natural
electroweak symmetry breaking suggests that superpartners should
already have been observed. The combination of accelerated unification
and twisted supersymmetry breaking can alleviate the tension between
these two facts, since the couplings unify at a much lower scale, and
supersymmetry breaking can arise even at tree-level in the case where
an extra dimension forms. In this case, accelerated
unification arises from scales above those where an extra
dimension forms, while supersymmetry breaking arises from beneath this
scale. The ability to address high energy scales is thus crucial.

Our excursions in theory space
\cite{Arkani-Hamed:2001ca,Arkani-Hamed:2001nc,Arkani-Hamed:2001vr}
have thus far been motivated by specific physical problems.  Twisted
supersymmetry breaking is yet another illustration of the unexpected
physics that can arise from the interplay of sites and links.  The
surprising ease with which theory space has provided new approaches to
problems as diverse as UV completion of higher-dimensional gauge
theories, stabilization of the electroweak scale, low scale gauge
coupling unification, and supersymmetry breaking, suggests to us
that there are deeper ideas surrounding these structures that we have
yet to fathom.

\begin{acknowledgments} 
  We would like the thank Ann Nelson and David Kaplan for stimulating
  discussions at the inception of this work.  We thank Savas
  Dimopoulos for discussions on combining twisted susy with
  accelerated unification. NA-H thanks Jay Wacker for topological
  tutelage.  H.G. is supported in part by the National Science
  Foundation under grant number NSF-PHY/98-02709. A.G.C. is supported
  in part by the Department of Energy under grant number
  \#DE-FG02-91ER-40676.  N.A-H.  is supported in part by the
  Department of Energy. under Contracts DE-AC03-76SF00098, the
  National Science Foundation under grant PHY-95-14797, the Alfred P.
  Sloan foundation, and the David and Lucille Packard Foundation.
\end{acknowledgments}


\begin{thebibliography}{25}
\expandafter\ifx\csname natexlab\endcsname\relax\def\natexlab#1{#1}\fi
\expandafter\ifx\csname bibnamefont\endcsname\relax
  \def\bibnamefont#1{#1}\fi
\expandafter\ifx\csname bibfnamefont\endcsname\relax
  \def\bibfnamefont#1{#1}\fi
\expandafter\ifx\csname citenamefont\endcsname\relax
  \def\citenamefont#1{#1}\fi
\expandafter\ifx\csname url\endcsname\relax
  \def\url#1{\texttt{#1}}\fi
\expandafter\ifx\csname urlprefix\endcsname\relax\def\urlprefix{URL }\fi
\providecommand{\bibinfo}[2]{#2}
\providecommand{\eprint}[2][]{\url{#2}}

\bibitem[{\citenamefont{Georgi}(1986)}]{Georgi:1986hf}
\bibinfo{author}{\bibfnamefont{H.}~\bibnamefont{Georgi}},
  \bibinfo{journal}{Nucl. Phys.} \textbf{\bibinfo{volume}{B266}},
  \bibinfo{pages}{274} (\bibinfo{year}{1986}).

\bibitem[{\citenamefont{Douglas and Moore}(1996)}]{Douglas:1996sw}
\bibinfo{author}{\bibfnamefont{M.~R.} \bibnamefont{Douglas}} \bibnamefont{and}
  \bibinfo{author}{\bibfnamefont{G.}~\bibnamefont{Moore}}
  (\bibinfo{year}{1996}), \eprint{hep-th/9603167}.

\bibitem[{\citenamefont{Kachru and Silverstein}(1998)}]{Kachru:1998ys}
\bibinfo{author}{\bibfnamefont{S.}~\bibnamefont{Kachru}} \bibnamefont{and}
  \bibinfo{author}{\bibfnamefont{E.}~\bibnamefont{Silverstein}},
  \bibinfo{journal}{Phys. Rev. Lett.} \textbf{\bibinfo{volume}{80}},
  \bibinfo{pages}{4855} (\bibinfo{year}{1998}), \eprint{hep-th/9802183}.

\bibitem[{\citenamefont{Bershadsky et~al.}(1998)\citenamefont{Bershadsky,
  Kakushadze, and Vafa}}]{Bershadsky:1998mb}
\bibinfo{author}{\bibfnamefont{M.}~\bibnamefont{Bershadsky}},
  \bibinfo{author}{\bibfnamefont{Z.}~\bibnamefont{Kakushadze}},
  \bibnamefont{and} \bibinfo{author}{\bibfnamefont{C.}~\bibnamefont{Vafa}},
  \bibinfo{journal}{Nucl. Phys.} \textbf{\bibinfo{volume}{B523}},
  \bibinfo{pages}{59} (\bibinfo{year}{1998}), \eprint{hep-th/9803076}.

\bibitem[{\citenamefont{Arkani-Hamed
  et~al.}(2001{\natexlab{a}})\citenamefont{Arkani-Hamed, Cohen, and
  Georgi}}]{Arkani-Hamed:2001ca}
\bibinfo{author}{\bibfnamefont{N.}~\bibnamefont{Arkani-Hamed}},
  \bibinfo{author}{\bibfnamefont{A.~G.} \bibnamefont{Cohen}}, \bibnamefont{and}
  \bibinfo{author}{\bibfnamefont{H.}~\bibnamefont{Georgi}},
  \bibinfo{journal}{Phys. Rev. Lett.} \textbf{\bibinfo{volume}{86}},
  \bibinfo{pages}{4757} (\bibinfo{year}{2001}{\natexlab{a}}),
  \eprint{hep-th/0104005}.

\bibitem[{\citenamefont{Hill et~al.}(2001)\citenamefont{Hill, Pokorski, and
  Wang}}]{Hill:2000mu}
\bibinfo{author}{\bibfnamefont{C.~T.} \bibnamefont{Hill}},
  \bibinfo{author}{\bibfnamefont{S.}~\bibnamefont{Pokorski}}, \bibnamefont{and}
  \bibinfo{author}{\bibfnamefont{J.}~\bibnamefont{Wang}}
  (\bibinfo{year}{2001}), \eprint{hep-th/0104035}.

\bibitem[{\citenamefont{Cheng et~al.}(2001{\natexlab{a}})\citenamefont{Cheng,
  Hill, Pokorski, and Wang}}]{Cheng:2001vd}
\bibinfo{author}{\bibfnamefont{H.-C.} \bibnamefont{Cheng}},
  \bibinfo{author}{\bibfnamefont{C.~T.} \bibnamefont{Hill}},
  \bibinfo{author}{\bibfnamefont{S.}~\bibnamefont{Pokorski}}, \bibnamefont{and}
  \bibinfo{author}{\bibfnamefont{J.}~\bibnamefont{Wang}}
  (\bibinfo{year}{2001}{\natexlab{a}}), \eprint{hep-th/0104179}.

\bibitem[{\citenamefont{Arkani-Hamed
  et~al.}(2001{\natexlab{b}})\citenamefont{Arkani-Hamed, Cohen, and
  Georgi}}]{Arkani-Hamed:2001nc}
\bibinfo{author}{\bibfnamefont{N.}~\bibnamefont{Arkani-Hamed}},
  \bibinfo{author}{\bibfnamefont{A.~G.} \bibnamefont{Cohen}}, \bibnamefont{and}
  \bibinfo{author}{\bibfnamefont{H.}~\bibnamefont{Georgi}},
  \bibinfo{journal}{Phys. Lett.} \textbf{\bibinfo{volume}{B513}},
  \bibinfo{pages}{232} (\bibinfo{year}{2001}{\natexlab{b}}),
  \eprint{hep-ph/0105239}.

\bibitem[{\citenamefont{Arkani-Hamed
  et~al.}(2001{\natexlab{c}})\citenamefont{Arkani-Hamed, Cohen, and
  Georgi}}]{Arkani-Hamed:2001vr}
\bibinfo{author}{\bibfnamefont{N.}~\bibnamefont{Arkani-Hamed}},
  \bibinfo{author}{\bibfnamefont{A.~G.} \bibnamefont{Cohen}}, \bibnamefont{and}
  \bibinfo{author}{\bibfnamefont{H.}~\bibnamefont{Georgi}}
  (\bibinfo{year}{2001}{\natexlab{c}}), \eprint{hep-th/0108089}.

\bibitem[{\citenamefont{Csaki et~al.}(2001{\natexlab{a}})\citenamefont{Csaki,
  Erlich, Grojean, and Kribs}}]{Csaki:2001em}
\bibinfo{author}{\bibfnamefont{C.}~\bibnamefont{Csaki}},
  \bibinfo{author}{\bibfnamefont{J.}~\bibnamefont{Erlich}},
  \bibinfo{author}{\bibfnamefont{C.}~\bibnamefont{Grojean}}, \bibnamefont{and}
  \bibinfo{author}{\bibfnamefont{G.~D.} \bibnamefont{Kribs}}
  (\bibinfo{year}{2001}{\natexlab{a}}), \eprint{hep-ph/0106044}.

\bibitem[{\citenamefont{Cheng et~al.}(2001{\natexlab{b}})\citenamefont{Cheng,
  Kaplan, Schmaltz, and Skiba}}]{Cheng:2001an}
\bibinfo{author}{\bibfnamefont{H.~C.} \bibnamefont{Cheng}},
  \bibinfo{author}{\bibfnamefont{D.~E.} \bibnamefont{Kaplan}},
  \bibinfo{author}{\bibfnamefont{M.}~\bibnamefont{Schmaltz}}, \bibnamefont{and}
  \bibinfo{author}{\bibfnamefont{W.}~\bibnamefont{Skiba}}
  (\bibinfo{year}{2001}{\natexlab{b}}), \eprint{hep-ph/0106098}.

\bibitem[{\citenamefont{Cheng et~al.}(2001{\natexlab{c}})\citenamefont{Cheng,
  Matchev, and Wang}}]{Cheng:2001qp}
\bibinfo{author}{\bibfnamefont{H.-C.} \bibnamefont{Cheng}},
  \bibinfo{author}{\bibfnamefont{K.~T.} \bibnamefont{Matchev}},
  \bibnamefont{and} \bibinfo{author}{\bibfnamefont{J.}~\bibnamefont{Wang}}
  (\bibinfo{year}{2001}{\natexlab{c}}), \eprint{hep-ph/0107268}.

\bibitem[{\citenamefont{Csaki et~al.}(2001{\natexlab{b}})\citenamefont{Csaki,
  Kribs, and Terning}}]{Csaki:2001qm}
\bibinfo{author}{\bibfnamefont{C.}~\bibnamefont{Csaki}},
  \bibinfo{author}{\bibfnamefont{G.~D.} \bibnamefont{Kribs}}, \bibnamefont{and}
  \bibinfo{author}{\bibfnamefont{J.}~\bibnamefont{Terning}}
  (\bibinfo{year}{2001}{\natexlab{b}}), \eprint{hep-ph/0107266}.

\bibitem[{\citenamefont{Cheng et~al.}(2001{\natexlab{d}})\citenamefont{Cheng,
  Hill, and Wang}}]{Cheng:2001nh}
\bibinfo{author}{\bibfnamefont{H.-C.} \bibnamefont{Cheng}},
  \bibinfo{author}{\bibfnamefont{C.~T.} \bibnamefont{Hill}}, \bibnamefont{and}
  \bibinfo{author}{\bibfnamefont{J.}~\bibnamefont{Wang}}
  (\bibinfo{year}{2001}{\natexlab{d}}), \eprint{hep-ph/0105323}.

\bibitem[{\citenamefont{He et~al.}(2001)\citenamefont{He, Hill, and
  Tait}}]{He:2001fz}
\bibinfo{author}{\bibfnamefont{H.-J.} \bibnamefont{He}},
  \bibinfo{author}{\bibfnamefont{C.~T.} \bibnamefont{Hill}}, \bibnamefont{and}
  \bibinfo{author}{\bibfnamefont{T.~M.~P.} \bibnamefont{Tait}}
  (\bibinfo{year}{2001}), \eprint{hep-ph/0108041}.

\bibitem[{\citenamefont{Bander}(2001)}]{Bander:2001qk}
\bibinfo{author}{\bibfnamefont{M.}~\bibnamefont{Bander}}
  (\bibinfo{year}{2001}), \eprint{hep-th/0107130}.

\bibitem[{\citenamefont{Sfetsos}(2001)}]{Sfetsos:2001qb}
\bibinfo{author}{\bibfnamefont{K.}~\bibnamefont{Sfetsos}}
  (\bibinfo{year}{2001}), \eprint{hep-th/0106126}.

\bibitem[{\citenamefont{Dai and Song}(2001)}]{Dai:2001bx}
\bibinfo{author}{\bibfnamefont{J.}~\bibnamefont{Dai}} \bibnamefont{and}
  \bibinfo{author}{\bibfnamefont{X.-C.} \bibnamefont{Song}}
  (\bibinfo{year}{2001}), \eprint{hep-ph/0105280}.

\bibitem[{\citenamefont{Alishahiha}(2001)}]{Alishahiha:2001nb}
\bibinfo{author}{\bibfnamefont{M.}~\bibnamefont{Alishahiha}}
  (\bibinfo{year}{2001}), \eprint{hep-th/0105153}.

\bibitem[{\citenamefont{Scherk and Schwarz}(1979)}]{Scherk:1979zr}
\bibinfo{author}{\bibfnamefont{J.}~\bibnamefont{Scherk}} \bibnamefont{and}
  \bibinfo{author}{\bibfnamefont{J.~H.} \bibnamefont{Schwarz}},
  \bibinfo{journal}{Nucl. Phys.} \textbf{\bibinfo{volume}{B153}},
  \bibinfo{pages}{61} (\bibinfo{year}{1979}).

\bibitem[{\citenamefont{Antoniadis et~al.}(1998)\citenamefont{Antoniadis,
  Dimopoulos, and Dvali}}]{Antoniadis:1998zg}
\bibinfo{author}{\bibfnamefont{I.}~\bibnamefont{Antoniadis}},
  \bibinfo{author}{\bibfnamefont{S.}~\bibnamefont{Dimopoulos}},
  \bibnamefont{and} \bibinfo{author}{\bibfnamefont{G.}~\bibnamefont{Dvali}},
  \bibinfo{journal}{Nucl. Phys.} \textbf{\bibinfo{volume}{B516}},
  \bibinfo{pages}{70} (\bibinfo{year}{1998}), \eprint{hep-ph/9710204}.

\bibitem[{\citenamefont{Barbieri et~al.}(2001)\citenamefont{Barbieri, Hall, and
  Nomura}}]{Barbieri:2000vh}
\bibinfo{author}{\bibfnamefont{R.}~\bibnamefont{Barbieri}},
  \bibinfo{author}{\bibfnamefont{L.~J.} \bibnamefont{Hall}}, \bibnamefont{and}
  \bibinfo{author}{\bibfnamefont{Y.}~\bibnamefont{Nomura}},
  \bibinfo{journal}{Phys. Rev.} \textbf{\bibinfo{volume}{D63}},
  \bibinfo{pages}{105007} (\bibinfo{year}{2001}), \eprint{hep-ph/0011311}.

\bibitem[{\citenamefont{Arkani-Hamed
  et~al.}(2001{\natexlab{d}})\citenamefont{Arkani-Hamed, Hall, Nomura, Smith,
  and Weiner}}]{Arkani-Hamed:2001mi}
\bibinfo{author}{\bibfnamefont{N.}~\bibnamefont{Arkani-Hamed}},
  \bibinfo{author}{\bibfnamefont{L.~J.} \bibnamefont{Hall}},
  \bibinfo{author}{\bibfnamefont{Y.}~\bibnamefont{Nomura}},
  \bibinfo{author}{\bibfnamefont{D.~R.} \bibnamefont{Smith}}, \bibnamefont{and}
  \bibinfo{author}{\bibfnamefont{N.}~\bibnamefont{Weiner}},
  \bibinfo{journal}{Nucl. Phys.} \textbf{\bibinfo{volume}{B605}},
  \bibinfo{pages}{81} (\bibinfo{year}{2001}{\natexlab{d}}),
  \eprint{hep-ph/0102090}.

\bibitem[{\citenamefont{Banks}(2000)}]{Banks:2000fe}
\bibinfo{author}{\bibfnamefont{T.}~\bibnamefont{Banks}} (\bibinfo{year}{2000}),
  \eprint{hep-th/0007146}.

\bibitem[{\citenamefont{Banks and Fischler}(2000)}]{Banks:2000gw}
\bibinfo{author}{\bibfnamefont{T.}~\bibnamefont{Banks}} \bibnamefont{and}
  \bibinfo{author}{\bibfnamefont{W.}~\bibnamefont{Fischler}}
  (\bibinfo{year}{2000}), \eprint{hep-th/0007186}.

\end{thebibliography}

\end{document}